\documentclass[11pt]{article}
\usepackage[utf8]{inputenc}
\usepackage{comment}
\usepackage{tikz}
\usepackage{ytableau}
\usepackage{xcolor}
\usepackage{graphicx}
\usepackage{enumitem}
\usepackage{multicol}
\usepackage{amsfonts}
\usepackage{mathrsfs}
\usepackage{wrapfig}
\usepackage{multicol}
\usepackage[top = 2 cm, bottom = 2 cm, left = 2.5 cm, right = 1.5 cm]{geometry}
\usepackage{amsmath,amsthm,amssymb}

\usepackage{cite,enumerate,float,indentfirst}

\usepackage{color}
\definecolor{red}{rgb}{1,0,0}

\graphicspath{{pictures/}}
\DeclareGraphicsExtensions{.pdf,.png,.jpg}

\begin{document}

\title{\vspace{1.5cm}\bf
$\beta$-WLZZ models from $\beta$-ensemble integrals directly
}

\author{
A. Mironov$^{b,c,d,}$\footnote{mironov@lpi.ru,mironov@itep.ru},
A. Oreshina$^{a,c,d,}$\footnote{oreshina.aa@phystech.edu},
A. Popolitov$^{a,c,d,}$\footnote{popolit@gmail.com}
}

\date{ }

\maketitle

\vspace{-6.5cm}

\begin{center}
\hfill FIAN/TD-06/24\\
\hfill IITP/TH-12/24\\
\hfill ITEP/TH-14/24\\
\hfill MIPT/TH-10/24
\end{center}

\vspace{4.5cm}

\begin{center}
$^a$ {\small {\it MIPT, Dolgoprudny, 141701, Russia}}\\
$^b$ {\small {\it Lebedev Physics Institute, Moscow 119991, Russia}}\\
$^c$ {\small {\it National Research Center ``Kurchatov Institute", 123182, Moscow, Russia}}\\
$^d$ {\small {\it Institute for Information Transmission Problems, Moscow 127994, Russia}}
\end{center}

\vspace{.1cm}

\begin{abstract}
Recently, we performed a two $\beta$-ensemble realization of the series of $\beta$-deformed WLZZ matrix models involving $\beta$-deformed Harish-Chandra-Itzykson-Zuber integrals. The realization was derived and studied by using Ward identities, which do not allow one to fix integration contours, these latter were chosen to be real axis for one $\beta$-ensemble and imaginary axis for the other one based on some particular checks. Here, we evaluate the $\beta$-ensemble integrals directly using a conjecture by I.G. Macdonald, and explain that another choice of integration contours is also possible.
\end{abstract}

\paragraph{1.} Matrix models have extremely many applications, from quantum theory of collisions \cite{Wigner} to theory of $2d$ gravity coupled to matter \cite{2dg1,2dg2,2dg3,2dg4} and Wilson averages in Chern-Simons theory \cite{CS1,CS2,CS3}. One of the basic applications of matrix models is related to their deformed versions, the so called $\beta$-ensembles \cite{Mehta,beta1,beta2,beta3,beta4}. The $\beta$-ensembles are associated with deformation of the invariant matrix model integral after performing integration over the angular variables. The Jacobian of the transformation from matrices to the eigenvalues plus angular variables is the square of the Vandermonde determinant, and coming to the $\beta$-ensemble implies dealing with an arbitrary degree of this determinant instead of 2. This determinant can be interpreted as a logarithmic potential in the $2d$ Coulomb system \cite{Dyson}.
A recent application also relates the $\beta$-ensembles to the AGT conjecture \cite{AGTmamo1,AGTmamo2,AGTmamo3,AGTmamo4,AGTmamo5}.

While the theory of $\beta$-ensembles is developed relatively far, models of a few interacting $\beta$-ensembles have not attracted too much attention so far: they were discussed only in \cite{Ey1} and in a recent paper by the authors \cite{MOP} (see also \cite{Ch2} for some preliminary discussion). At the same time, they correspond to a $\beta$-deformed version of two-matrix models studied rather intensively \cite{Gava,MMM1,MMM2,AMMN,Al,Ch1,Ch2}. The reason is that the models involving a few $\beta$-ensembles require knowledge of $\beta$-deformation \cite{BH,Eynard,MMS} of the Harish-Chandra-Itzykson-Zuber (HCIZ) integral \cite{HC,IZ}, which is quite a non-trivial problem, so far lacking a complete solution: there are only some recurrent formulas applicable at natural values of $\beta$ \cite{BH,Eynard}, a definition in terms of formal series \cite{MMS} and a set of identities for the integrals \cite{MOP}. At last, there are identities that are known under the name of Macdonald conjecture \cite{Mac,Rosler}, which we are going to use in this note.

The central character of the play is a $W$-representation for the matrix models and for their $\beta$-ensemble versions \cite{China121,China122,MO} that leads to the partition functions of the form
        \begin{equation}\label{Z1Jack}
            Z^{(\beta)}(N;p,g)=\sum_{R} \xi_R^\beta(N) {J_{R}\{ p_k\}J_R\{g_k\} \over ||J_R||}
        \end{equation}
where $J_R$ are the Jack polynomials \cite{Mac2}, the summation goes over partitions $R$, and
\begin{equation}\label{xi}
            \xi_R^\beta(N):=\prod_{i,j\in R}(N+\beta^{-1}(j-1)-i+1)=\beta^{-|R|}\prod_{j=1}^{N}\frac{\Gamma(\beta(N-j+1)+R_j)}{\Gamma(\beta(N-j+1))}
        \end{equation}
Such models are often referred to as $\beta$-WLZZ models, in accordance with their non-deformed counterpart, the WLZZ models first constructed in \cite{China121,China122}.

In the previous paper \cite{MOP}, we constructed an integral representation for the $\beta$-WLZZ class of matrix models:
    \begin{equation} \label{PartFuncLambda}
Z^{(\beta)}(N;p,g)=\int[dxdy]  \Delta^{2\beta} (x) \Delta^{2\beta} (y) I^{\beta}(x,-y)  \exp \left[\beta \sum_{k \geq 1} \frac{g_k}{k} \sum_{j=1}^{N} x_j^k + \beta \sum_{k \geq 1} \frac{\bar p_k}{k}\sum_{j=1}^{N} y_j^{{ k}} \right]
    \end{equation}
where the integrals over $x_j$ run over the real axis, and integrals over $y_j$ run over the imaginary one. Here $I^{\beta}(x,y)$ is the $\beta$-deformed HCIZ ($\beta$-HCIZ) integral \cite{BH,Eynard}, and the measure is normalized in such a way that $Z_\beta(N;0,0,0)=1$.
Here $\Delta(x)=\prod_{i<j}|x_i-x_j|$ is the Vandermonde determinant, and the integrand is understood as a formal series in $p_k$ and $g_k$.

In \cite{MOP}, we proved that the partition functions (\ref{PartFuncLambda}) and (\ref{Z1Jack}) are equal to each other by using Ward identities which they satisfy. However, this kind of check does not fix integration contours, and we checked it in some special cases at natural values of $\beta$. The goal of this note is to derive this equality directly.

\paragraph{2.} Throughout the note, we define the norm of Jack polynomials to be
\begin{equation}\label{norm}
||J_R||:={\overline{G}^\beta_{R^\vee R}(0)\over G^\beta_{RR^\vee}(0)}\beta^{|R|}\ \ \ \ \ \ \
G_{R'R''}^\beta(x):=\prod_{(i,j)\in R'}\Big(x+R'_i-j+\beta(R''_j- i+1)\Big)
\end{equation}
with the bar over the functions denoting the substitution $\beta\to\beta^{-1}$.

Now, first of all, using the Cauchy identity \cite{Mac2}
        \begin{equation}
            \exp \left[ \beta\sum_{k \geq 1} \frac{g_k}{k} \sum_{j=1}^{N} x_j^k \right]=\sum_{R} \frac{J_{R}(x) J_{R} (g_k)}{||J_R||}
        \end{equation}
        one can obtain for the integral (\ref{PartFuncLambda}) (for any integration contour):
        \begin{multline}\label{int}
            Z^{(\beta)}(N;p,g)
            =\sum_{R,Q}\frac{J_R(p_k)J_{Q}(g_k)}{||J_R|| ||J_Q||} \int[dy] \int[dx]  \Delta^{2\beta} (x) \Delta^{2\beta} (y) I^{\beta}(x,-y) J_R(Y) J_{Q}(X)=\\
            =\sum_{R,Q}\frac{J_R(p_k)J_{Q}(g_k)}{||J_R|| ||J_Q||} < J_R(Y) J_{Q}({ X})>
        \end{multline}
Thus, in order to prove the equivalence of (\ref{PartFuncLambda}) and (\ref{Z1Jack}), it suffices to derive that
\begin{equation}\label{be}
< J_R(Y) J_{Q}({ X})> = \delta_{R,Q} \xi_R^\beta(N)  ||J_R||
\end{equation}

Earlier \cite{MOP}, in order to evaluate integral \eqref{int} at natural $\beta$, we used the Fourier theory formula
        \begin{equation}
            \left.\int dxdyf(x)g(y)e^{-xy} = f \left( {\partial \over \partial x} \right) g(x) \right|_{x=0}
        \end{equation}
and also the $\beta$-HCIZ integral at natural $\beta$ in the form \cite{Eynard}:
        \begin{equation}
            I^\beta_N(x,y)= \sum_{\sigma} \frac{e^{\sum_{j=1}^{N} x_j y_{\sigma(j)}}}{\Delta(x)^{2\beta}\Delta(y_{\sigma})^{2\beta}} \Tilde{I}_N^{\beta} (x,y_{\sigma})
        \end{equation}
where the sum runs over permutations of $y_i$, and $\Tilde{I}_N^{\beta}$ is a polynomial of $x_i$ and $y_i$ defined by recurrent relations in $N$.

Now, we consider scalar products with two different integration contours,
\begin{equation} \label{aver1}
            < f(X,Y) >_1:=\int_{- i \infty}^{+i \infty}[dy] \int_{- \infty}^{+ \infty}[dx] \Delta^{2\beta} (x) \Delta^{2\beta} (y) I^{\beta}(x,-y) f(X,Y)
        \end{equation}
and
\begin{equation} \label{aver2}
            < f(X,Y) >_2:=\oint_{0}[dy] \int_{0}^{+ \infty}[dx]  \Delta^{2\beta} (x) \Delta^{2\beta} (y) I^{\beta}(x,-y) f(X,Y)
        \end{equation}
derive (\ref{be}) for the second one, and then demonstrate the equivalence of the scalar products.

\paragraph{3.} Let us define the $\beta$-HCIZ integral at arbitrary $\beta$ as a power series  \cite{MMS}:
\begin{equation}\label{IZJack}
        I_N^\beta(x,y) = \sum_{R} {1\over \xi_R^{(\beta)}(N)}{J_R(x) J_R(y)\over||J_R||}
    \end{equation}
Then, in order to prove (\ref{be}) with the second scalar product (\ref{aver1}), we use the following facts:
        \begin{itemize}
            \item The Macdonald conjecture that claims \cite{Mac,Rosler}:
                \begin{equation}\label{MC}
                    \int_0^{\infty}I^\beta_N(x,-y)J_R(x)\Delta(x)^{2\beta}\prod_{j=1}^Nx_j^{\mu-\beta(N-1)-1}dx_j=N_R\cdot J_R(y^{-1})\prod_{j=1}^Ny_j^{-\mu}
                \end{equation}
                \begin{equation}
                    N_R=\prod_{j=1}^N{\Gamma(1+\beta j)\over\Gamma (1+\beta)}\Gamma(R_j+\mu-\beta(j-1))
                \end{equation}
            \item
                The Jack polynomials orthogonality \cite[(10.38)]{Mac2} w.r.t. the scalar product
                \begin{equation}
                    \Big<f,g\Big>:=\oint_0\prod_{j=1}^N{dy_j\over y_j}f(y^{-1})g(y)\prod_{j\ne k}\Big(1-{y_j\over y_k}\Big)^\beta
                \end{equation}
                so that
                \begin{equation}
                    \Big<J_R,J_Q\Big>={\bf N}_R\cdot\delta_{RQ}
                \end{equation}
                \begin{equation}
                    {\bf N}_R=N! \prod_{1 \leq i<j\leq N} \frac{\Gamma(R_i-R_j+ \beta (j-i+1))\Gamma(R_i-R_j+ \beta (j-i-1)+1)}{\Gamma(R_i-R_j+ \beta (j-i))\Gamma(R_i-R_j+ \beta (j-i)+1)}
                \end{equation}
        \end{itemize}
        Integrating now the both sides of equation (\ref{MC}) with the measure $\Delta(y)^{2\beta}$ and choosing $\mu=\beta(N-1)+1$, the second scalar product gives rise to
        \begin{equation}
            < J_R(Y) J_{Q}({ X})>_2 = \delta_{R,Q}{ {\bf N}_R N_R \over {\bf N}_{\varnothing} N_{\varnothing}}=
             \delta_{R,Q}\xi_R^\beta(N)  ||J_R||
        \end{equation}
where we used (\ref{norm}) and (\ref{xi}).

\paragraph{4.} Now it remains to prove that the two scalar products coincide. Consider first the simplest example of $\beta=1$, $N=1$.
Then, the two integrals (\ref{aver1}) and (\ref{aver2}) over the two different contours give rise to the same result:
        \begin{equation}\label{I111}
            \int_{- i \infty}^{+i \infty}dy \int_{- \infty}^{+ \infty}dx e^{-xy} x^a y^b = a! \delta_{a,b}
        \end{equation}
        \begin{equation}\label{I112}
            \oint_{0} dy \int_{0}^{+ \infty} dx e^{-xy} x^a y^b = \oint_{0} dy \ y^b \frac{a!}{y^{1+a}} = a! \delta_{a,b}
        \end{equation}
However, the proof that generalizes to the $\beta\ne 1$ case should not use an explicit integration, hence, we establish now the equivalence of \eqref{I111} and \eqref{I112} without direct calculations. Indeed, note that
        \begin{equation}
            \int_{- i \infty}^{+i \infty}dy \int_{- \infty}^{+ \infty}dx e^{-xy} x^a y^b =\left\{
            \begin{array}{cl}
                2\int_{- i \infty}^{+i \infty}dy \int_{0}^{+ \infty}dx e^{-xy} x^a y^b  & \hbox{for even }a+b \\
                0 & \mbox{otherwise}.
            \end{array}
            \right.
        \end{equation}
Now, according to the Sokhotski theorem,
        \begin{equation}
            \oint_{0} dy \int_{0}^{+ \infty} dx e^{-xy} x^a y^b= \int_{- i \infty}^{+i \infty}dy \int_{0}^{+ \infty}dx e^{-xy} x^a y^b
        \end{equation}
This allows one to deform the contour in order to prove equivalence of the two scalar products, and this is simple to do in the generic case of arbitrary $\beta$ and $N$.

Indeed, notice that the polynomial
        \begin{equation}
            \mu(x,y)=  \Delta^{2\beta} (x) \Delta^{2\beta} (y) I^{\beta}(x,-y)
        \end{equation}
is invariant under simultaneous changing the sign of all variables: $x_i \rightarrow -x_i$, $y_i \rightarrow -y_i$. Thus
        \begin{equation}
            \int_{- i \infty}^{+i \infty}[dy] \int_{- \infty}^{+ \infty}[dx] \mu(x,y) \Bar{x}^{\Bar{a}} \Bar{y}^{\Bar{b}} =\left\{
            \begin{array}{cl}
                2\int_{- i \infty}^{+i \infty}[dy] \int_{0}^{+ \infty}[dx] \mu(x,y) \Bar{x}^{ \Bar{a} } \Bar{y} ^{ \Bar{b} } &
                \hbox{for even }a+b\\
                0 & \mbox{otherwise}.
            \end{array}
            \right.
        \end{equation}
where $\Bar{x}^{ \Bar{a} }=\prod_{i=1}^{N}x_i^{a_i}$. Finally, using the Sokhotski theorem, one obtains
        \begin{equation}
            \int_{- i \infty}^{+i \infty}[dy] \int_{- \infty}^{+ \infty}[dx] \mu(x,y) \Bar{x}^{\Bar{a}} \Bar{y}^{\Bar{b}}=\oint_{0}[dy] \int_{0}^{+ \infty}[dx] \mu(x,y) \Bar{x}^{\Bar{a}} \Bar{y}^{\Bar{b}}
        \end{equation}
which proves the equivalence of the two scalar products and immediately leads to
        \begin{multline}
             \int_{- i \infty}^{+i \infty}[dy] \int_{- \infty}^{+ \infty}[dx]  \Delta^{2\beta} (x) \Delta^{2\beta} (y) I^{\beta}(x,-y) \exp \left[ \beta\sum_{k \geq 1} \frac{g_k}{k} \sum_{j=1}^{N} x_j^k + \beta \sum_{k \geq 1} \frac{ p_k}{k}\sum_{j=1}^{N} y_j^m \right]=\\
             =\oint_{0}[dy] \int_{0}^{+ \infty}[dx]  \Delta^{2\beta} (x) \Delta^{2\beta} (y) I^{\beta}(x,-y) \exp \left[\beta \sum_{k \geq 1} \frac{g_k}{k} \sum_{j=1}^{N} x_j^k +  \beta\sum_{k \geq 1} \frac{ p_k}{k}\sum_{j=1}^{N} y_j^{{ k}} \right]
        \end{multline}
thus proving equivalence of (\ref{Z1Jack}) and (\ref{PartFuncLambda}).

\paragraph{CONFLICT OF INTEREST.}
The authors of this work declare that they have no conflicts of interest.

\paragraph{FUNDING.} Our work is partly supported by the grant of the Foundation for the Advancement of Theoretical Physics ``BASIS" (A. Mironov and A. Oreshina) and by the joint grant 21-51-46010-ST-a.

\end{document}